\documentclass[10pt]{iopart}
\usepackage{iopams}

\usepackage{color} \usepackage{graphicx}

\newcommand \be {\begin{equation}} \newcommand \ee {\end{equation}}
\newcommand \bea {\begin{eqnarray}} \newcommand \eea {\end{eqnarray}}
\newcommand \rhot {\tilde{\rho}} \newcommand \la {\langle} \newcommand
\ra {\rangle} \newcommand \ve {\varepsilon} \newcommand \ut
{\tilde{u}} 
\newcommand \Pe {\mathrm{Pe}}

\begin{document}

\title[Biased motility-induced phase separation]{Biased motility-induced phase separation: from chemotaxis to
traffic jams} \date{\today} \author{Eric Bertin$^1$,
Alexandre Solon$^2$}

\address{$1$ Universit\'e~Grenoble Alpes, CNRS, LIPhy, 38000
Grenoble, France}

\address{$2$ Sorbonne Université, CNRS, Laboratoire de Physique
Théorique de la Matière Condensée, 75005 Paris, France}

\begin{abstract} We propose a one-dimensional model of active
  particles interpolating between quorum sensing models used in the
  study of motility-induced phase separation (MIPS) and models of
  congestion of traffic flow on a single-lane highway. Particles have
  a target velocity with a density-dependent magnitude and a direction
  that flips with a finite rate that is biased toward moving
  right. Two key parameters are the bias and the speed
  relaxation time. MIPS is known to occur in such models at zero bias
  and zero relaxation time (overdamped dynamics), while a fully biased
  motion with no velocity reversal models traffic flow on a
  highway. Using both numerical simulations and continuum equations
  derived from the microscopic dynamics, we show that a single
  phase-separated state extends from the usual MIPS to congested
  traffic flow in the phase diagram defined by the bias and the speed
  relaxation time. However, in the fully biased case, inertia is
  essential to observe phase separation, making MIPS and congested
  traffic flow seemingly different phenomena if not simultaneously
  considering inertia and tumbling. We characterize the velocity of
  the dense phase, which is static for usual MIPS and moves backwards
  in traffic congestion. We also find that in presence of bias, the
  phase diagram becomes richer, with an additional transition between
  phase separation and a microphase separation that is seen above a
  threshold bias or relaxation rate.
\end{abstract}

\section{Introduction}
\label{sec:introduction} We are all painfully aware that an excessive
density of cars on a road leads to traffic jams. They are often
triggered at special points, for example by an entrance or by road
works, but can also happen spontaneously in a dense homogeneous car
flow, without any apparent external perturbation. These so-called
``phantom jams'' have been observed in the form a single large
jam~\cite{kerner_experimental_1997} or multiple small jams leading to
stop-and-go traffic~\cite{kerner_experimental_1998}. In both cases,
the jams move upstream at a constant speed, with a constant outflow of
cars from the jams~\cite{kerner_experimental_1996}. An important tool
to understand the formation of traffic jams is the ``fundamental
diagram'' that gives the mean car flow $J(\rho)$ as a function of car
density $\rho$. Empirical measurements show that it is
non-monotonous~\cite{hall_empirical_1986,neubert_single-vehicle_1999}:
there is an optimal density above which the speed $u(\rho)$ of cars
decreases faster than $1/\rho$ so that the flux $J=\rho u$ decreases
overall.

The decrease in the speed of cars with increasing density is
reminiscent of the behaviour of self-propelled particles undergoing
motility-induced phase separation (MIPS). Indeed, MIPS was first
predicted and observed numerically for collections of run-and-tumble
particles moving at a density-dependent speed
$u(\rho)$~\cite{tailleur_statistical_2008,cates_motility-induced_2015}. If
the decrease in speed is steep enough, precisely if $u'<-u/\rho$
(where $u'$ is the derivative of $u$ with respect to $\rho$), a
homogeneous system is unstable and undergoes MIPS: it phase separates
between a dense phase of slowly moving particles and a dilute phase of
fast movers (see Ref.~\cite{obyrne_introduction_2023} for a
pedagogical introduction). Interactions leading to a $u(\rho)$ are
naturally encountered in bacteria interacting via quorum-sensing
mediated by molecules that diffuse in the extracellular
medium~\cite{fu_stripe_2012,liu_self-driven_2019,curatolo_cooperative_2020},
in larger animals interacting via pheromones like
ants~\cite{anderson_social_2022} and can also be implemented in
colloidal systems~\cite{bauerle_self-organization_2018}. In addition,
although it does not capture the full
phenomenology~\cite{bialke_negative_2015,solon_generalized_2018-1,solon_generalized_2018,shi_self-organized_2020},
a function $u(\rho)$ is also a first approximation of the effect of
collisions that tend to slow down self-propelled disks interacting by
steric repulsion~\cite{speck_effective_2014}.  At high enough activity
and density, these systems also undergo MIPS as seen in
simulations~\cite{fily_athermal_2012,redner_structure_2013,stenhammar_continuum_2013}
and experiments with self-propelled
colloids~\cite{buttinoni_dynamical_2013,van_der_linden_interrupted_2019}.

The ingredients leading to MIPS and traffic jams appear to be similar
but the two phenomena have important differences. Perhaps the main one
is that roads (or at least lanes) are unidirectional whereas MIPS has
only been studied, to the best of our knowledge, in isotropic
systems. Another difference is in the importance of inertia: it is
thought to be crucial to describe the phenomenology of traffic flow
and is included in most models (see
Ref.~\cite{chowdhury_statistical_2000} for a review of many models)
whereas it is inessential to MIPS. On the contrary, inertia tends to
suppress MIPS for self-propelled
dumbbells~\cite{suma_motility-induced_2014} or
disks~\cite{mandal_motility-induced_2019} interacting via pairwise
repulsion. The effect of inertia on quorum-sensing particles
interacting via a speed $u(\rho)$ has not been assessed.

In this paper, we attempt to bridge the gap between the two
phenomena. To this end, we build on a one-dimensional quorum-sensing
model of MIPS in which run-and-tumble particles interact via a
density-dependent speed
$u(\rho)$~\cite{tailleur_statistical_2008,solon_active_2015,solon_generalized_2018-1,solon_generalized_2018}
by adding (i) an external bias on the tumble rate so that the
particles move preferentially to the right and (ii) inertia so that
the velocity relaxes to its preferred value in a finite time. For
pedagogical reasons, we introduce these ingredients one at a time and
first consider in Sec.~\ref{sec:MIPS} an overdamped biased model in
which left-moving particles tumble more frequently than right-moving
ones, with a bias parameter $b$ ranging from $0$ when the motion is
isotropic to $1$ when particles move only to the right. This is the
type of bias that has been observed, for example, for {\it E. Coli}
bacteria in presence of chemoattractant~\cite{berg_chemotaxis_1972}.
In this overdamped model, we observe MIPS with a dense phase that is
moving upstream, as traffic jams do. Although intermediate values of
the bias tend to favor phase separation, at larger bias values
$b\lesssim 1$, the phase separation disappears. In the fully biased
$b=1$ case that is closest to a traffic flow model, homogeneous
systems are stable at any level of activity and density so that phase
separation is prevented. In Sec.~\ref{sec:traffic}, we consider the
fully biased model of Sec.~\ref{sec:MIPS} with inertia. This can be
seen as a traffic flow model at a mesoscopic scale, intermediate
between microscopic models in the form of asymmetric exclusion
processes~\cite{schreckenberg_discrete_1995,evans_bose-einstein_1996,evans_exact_1999}
and phenomenological (deterministic) continuum equations
\cite{aw_resurrection_2000,siebel_fundamental_2006}. We find that if
the inertial time is large enough, the homogeneous state becomes again
unstable. Interestingly, depending on the inertial time, one observes
either a phase separation or a micro-phase separation. The transition
between the two types of patterns is similar to what is reported for
the convective Cahn-Hilliard model~\cite{golovin_convective_2001} and
happens at the point where a binodal line crosses a spinodal
line. Finally, in Sec.~\ref{sec:bridge} we consider the general case
of run-and-tumble particles with arbitrary bias $0\le b\le 1$ and
inertia. We show that one can continuously change the parameters to
interpolate, retaining a phase separation, between the classic
description of MIPS (isotropic motion with overdamped dynamics) and
the traffic flow regime (fully biased motion with underdamped
dynamics).

In each of the three sections, we first define the microscopic model
and derive the continuum equation giving the evolution of the density
field on macroscopic time and length scales. We then compute the phase
diagram based on a linear stability analysis and numerical integration
of the continuum equations and compare the predictions to numerical
simulations of the microscopic model.

%%%%%%%%%%%%%%%%%%%%%%%%%%%%%%%%%%%%%%%%%%%%%%%%%%%%%%%%%%%%%%%%%%%%%%%%%%%%%%%%%%%%%%%%%

\section{MIPS-like scenario: overdamped biased dynamics with tumbles}
\label{sec:MIPS}

\subsection{Definition of the model}

We consider a model of run-and-tumble particles in one dimension with
overdamped dynamics and biased tumbling rate.  The position $x_i(t)$
of a particle evolves according to the overdamped Langevin dynamics
\begin{equation}
  \label{eq:modelOD} \frac{dx_i}{dt} = u(\rhot)\, e_i(t) + \sqrt{2D}
\, \xi_i(t)
\end{equation} where $D$ is the (passive) diffusion coefficient and
$u(\rhot)$ the speed of the particle (assumed not to depend on $i$) in
the presence of a locally averaged density $\rhot$ defined as
\begin{equation}
  \label{eq:rho-kernel} \rhot(x_i) = \sum_j K(x_i-x_j)
\end{equation} with an averaging kernel $K(x)$ and where the sum on
$j$ runs on all particles, including particle $i$. The quantity
$e_i(t)=\pm 1$ indicates the direction of motion of particle $i$ at
time $t$, and it randomly switches (tumbles) with rates $\alpha$ for
the transition $-1 \to 1$, and with rate $\alpha (1-b)$ for the
transition $1\to -1$, where $0\le b \le 1$ quantifies the bias ($b=0$:
no bias; $b=1$: fully biased dynamics).  The white noise $\xi_i(t)$
has correlation
\begin{equation} \label{eq:white:noise} \la \xi_i(t) \xi_j(t')\ra =
\delta(t-t')\delta_{ij}.
\end{equation}

In numerical simulations, we will consider in all the paper a speed
that decreases linearly with density from $u_0$ in free space to $0$
when the density is greater than a critical value $\rho^*$:
\begin{equation}
  \label{eq:u-simu} u(\rho)=u_0\left(1-\frac{\rho}{\rho^*}\right)
\,\,\, \rm{if}\, \rho<\rho^*; \qquad u(\rho)=0\,\,\, \rm{otherwise}.
\end{equation} This form has been measured to be a good approximation
of the slowdown due to collisions in (unbiased) self-propelled
repulsive disks~\cite{bialke_microscopic_2013}, and even becomes an
exact expression in the limit of infinite
dimensions~\cite{de_pirey_active_2019}. Moreover, it also
qualitatively reproduces the fundamental diagram of traffic
flow~\cite{siebel_fundamental_2006} with a parabolic profile for the
flux $J=\rho u(\rho)$ as a function of density, as in the asymmetric
exclusion process. Note that the vanishing velocity when $\rho>\rho^*$
does not lead to a
condensation~\cite{golestanian_bose-einstein-like_2019,mahault_boseeinstein-like_2020}
because of the non-zero positional diffusion $D$. We thus expect our
results to be robust to the addition of a small non-zero velocity when
$\rho>\rho^*$.  To compute the local averaged density in
Eq.~(\ref{eq:rho-kernel}), we use the bell-shaped kernel
\begin{equation}
  \label{eq:kernel}
K(x)=\frac{1}{Z}\exp\left(-\frac{1}{r_0^2-x^2}\right)\,\,\, {\rm if}\,
|x|<r_0; \quad\; K(x)=0\,\,\, {\rm otherwise},
\end{equation} with the normalization constant $Z$ such that
$\int_{-\infty}^{+\infty}K(x)dx=1$. It has a finite support
$[-r_0;r_0]$. In the following we fix the unit of length by choosing
$r_0=1$.

In absence of bias ($b=0$), the model reduces to the 1d version of the
quorum-sensing active particles (QSAPs) studied in
Ref.~\cite{tailleur_statistical_2008,solon_active_2015,solon_generalized_2018-1,solon_generalized_2018}.

\subsection{Hydrodynamic description}

We introduce the conditional densities $\rho_{+}(x,t)$ and
$\rho_{-}(x,t)$ of particles moving to the right and to the left
respectively. The total density is given by
$\rho(x,t)=\rho_{+}(x,t)+\rho_{-}(x,t)$. Making the local mean-field
approximation that $\rho_{\pm}(x,t)$ are not correlated to the
averaged local density $\rhot(x,t)$, the evolution of $\rho_{+}$ and
$\rho_{-}$ is governed by
\begin{eqnarray}
 \label{eq:rhop:ovdp} \partial_t \rho_{+} + \partial_x [u(\rhot)
\rho_{+}] &=& D \partial_x^2 \rho_{+} - \alpha(1-b) \rho_{+} + \alpha
\rho_{-},\\
 \label{eq:rhom:ovdp} \partial_t \rho_{-} - \partial_x [u(\rhot)
\rho_{-}] &=& D \partial_x^2 \rho_{+} - \alpha \rho_{-} + \alpha(1-b)
\rho_{+}.
\end{eqnarray}
Throughout the paper, we use the shorthand notation
$\partial_z^n$ to denote the $n^{\mathrm{th}}$-derivative with respect
to a variable $z$.  The total density $\rho$ is a conserved quantity,
whose evolution follows
\begin{equation} \label{eq:rho:ovdp:v0} \partial_t \rho + \partial_x
\left[u(\rhot) \big(\rho_{+}-\rho_{-}\big)\right] = D \partial_x^2
\rho.
\end{equation} To close the equation, we need to
reexpress $\rho_{+}-\rho_{-}$ in terms of $\rho$ in
Eq.~(\ref{eq:rho:ovdp:v0}).  This is done by identifying a relevant
fast variable that can be eliminated. One might try to use
$\rho_{+}-\rho_{-}$ as a fast variable, as was done in the unbiased
case~\cite{tailleur_statistical_2008}. However, in the limit of large bias, $\rho_{+}-\rho_{-}\approx \rho_{+}\approx \rho$ which is not a fast variable.
We find that an appropriate quantity to consider needs to remain a fast variable for all values of the bias and to vanish in a steady-state spatially homogeneous system so
that it can be expressed as a gradient of a function of
the density $\rho$ after an appropriate coarse-graining in time. A
natural choice is then to take as fast variable a field proportional
to the probability current between configurations $e_i=1$ and
$e_i=-1$, and we thus define the field $q(x,t)$ as
\begin{equation}
q = \rho_{-} - (1-b) \rho_{+}.
\end{equation}
The quantity $q$ indeed vanishes in a steady-state
homogeneous system. From Eqs.~(\ref{eq:rhop:ovdp}) and
(\ref{eq:rhom:ovdp}), we get
\begin{equation} \label{eq:g:ovdp:v0} \partial_t q - \partial_x
\left[u(\rhot) \big(\rho_{-}+(1-b)\rho_{+}\big)\right] = D
\partial_x^2 q - \alpha (2-b) q.
\end{equation}
Defining
\begin{equation} \ve = \frac{b}{2-b},
\end{equation} we get from Eq.~(\ref{eq:rho:ovdp:v0})
\begin{eqnarray}
  \label{eq:rho:ovdp:v1} \partial_t \rho + \partial_x \left[\ve \ut \rho -(1+\ve) \ut q \right] &=& D \partial_x^2 \rho,\\
\label{eq:g:ovdp:v1} \partial_t q - \partial_x \left[(1+\ve)(1-b) \ut \rho + \ve \ut q \right] &=& D \partial_x^2 q - \alpha (2-b) q,
\end{eqnarray} with $\ut\equiv u(\rhot)$.  On time scales much larger
that $\left[\alpha(2-b)\right]^{-1}$, the time derivative $\partial_t q$ may in practice
be neglected, and we obtain from Eq.~(\ref{eq:g:ovdp:v1}) an explicit
expression of $q$.  As we aim to expand the evolution equation
(\ref{eq:rho:ovdp:v1}) for $\rho$ to second order in gradient
(drift-diffusion order), we need to express $q$ only up to first order
in gradients, since $q$ appears within a gradient in
Eq.~(\ref{eq:rho:ovdp:v1}).  After truncation, we get
\begin{equation} \label{eq:g:deriv:urho} q = \frac{2(1-b)}{\alpha
(2-b)^2} \, \partial_x (\ut\rho).
\end{equation} Using Eq.~(\ref{eq:g:deriv:urho}) in
Eq.~(\ref{eq:rho:ovdp:v1}), we thus eventually obtain the evolution
equation for $\rho$, at drift-diffusion order,
\begin{equation} \label{eq:evolrho:final:mips} \partial_t \rho =
-\partial_x \big(\ve \rho \ut - \eta_0 \rho \ut \partial_x \ut\big) +
\partial_x \left[ \big( \eta_0 \ut^2 + D \big) \partial_x \rho \right],
\end{equation} where
\begin{equation} \label{eq:def:eta0} \eta_0 = \frac{4(1-b)}{\alpha
(2-b)^3}.
\end{equation}
To get a fully explicit evolution equation for $\rho$, we need to
express $\ut \equiv u(\rhot)$ in terms of $\rho$ and its spatial
derivatives.  Expanding $\rhot=\rho+\kappa \partial_x^2\rho$ as
in~\cite{solon_generalized_2018}, we get
\begin{equation}
  \label{eq:ut-exp}\ut \equiv u(\rhot) = u + \kappa u'
\partial_x^2\rho,
\end{equation} where the prime stands for a derivative with respect to
$\rho$.  Eq.~(\ref{eq:evolrho:final:mips}) then takes the form
\begin{equation} \partial_t \rho = -\partial_x J
\end{equation} with a particle current $J$ given by
\begin{equation}
  \label{eq:Jmips} J = J_0(\rho) - D_{\mathrm{eff}}(\rho) \partial_x
\rho + m(\rho) \partial_x^2 \rho + \Gamma_1(\rho) \partial_x^3 \rho -
\Gamma_2(\rho) \partial_x \rho \,\partial_x^2 \rho,
\end{equation} where
\begin{eqnarray}
\label{eq:def:J0:mips} J_0(\rho) &=& \ve\rho u,\\
\label{eq:def:Deff:mips} D_{\mathrm{eff}}(\rho) &=& D+ \eta_0 u (\rho
u)',\\ m(\rho) &=& \kappa \ve \rho u',\\
\label{eq:def:Gamma1:mips} \Gamma_1(\rho) &=& -\eta_0 \kappa \rho u
u',\\ \Gamma_2(\rho) &=& \eta_0 \kappa [\rho (uu''+u'^2)+2uu'].
\end{eqnarray}

\subsection{Linear stability analysis}
\label{sec:linear-stabl-OD}

We linearize Eq.~(\ref{eq:evolrho:final:mips}) around the homogeneous
state of density $\rho=\rho_0$, setting
$\rho(x,t)=\rho_0+\delta\rho(x,t)$, which yields to first order in
$\delta \rho$
\begin{equation} \label{eq:evol:rho:lin:mips} \partial_t \delta \rho =
- J_0'(\rho_0) \partial_x \delta \rho + D_{\mathrm{eff}}(\rho_0)
\partial_x^2 \delta \rho - m(\rho_0) \partial_x^3 \delta \rho -
\Gamma_1(\rho_0) \partial_x^4 \delta \rho.
\end{equation}
The eigenmodes of Eq.~(\ref{eq:evol:rho:lin:mips}) are Fourier modes
of the form
\begin{equation} \delta \rho(x) = \delta\rho_0 \, e^{st+iqx},
\end{equation} with a (complex) growth rate $s$ given by
\begin{equation} s = - iq J_0'(\rho_0) - q^2 D_{\mathrm{eff}}(\rho_0)
+ i q^3 m(\rho_0) - q^4 \Gamma_1(\rho_0).
\end{equation} An instability occurs when ${\rm Re}(s) \ge 0$, with
\begin{equation} {\rm Re}(s) = -q^2 D_{\mathrm{eff}}(\rho_0)- q^4
\Gamma_1(\rho_0).
\end{equation}
According to the expressions (\ref{eq:def:Deff:mips}) and
(\ref{eq:def:Gamma1:mips}) of $D_{\mathrm{eff}}$ and $\Gamma_1$
respectively, if $u(\rho)$ decreases fast enough as a function of
$\rho$, $D_{\mathrm{eff}}(\rho_0)$ may become negative, while
$\Gamma_1(\rho_0) >0$. In this case, an instability of the homogeneous
state occurs for wavenumbers $q<q^*$, with $q^* =
\sqrt{|D_{\mathrm{eff}}(\rho_0)|/\Gamma_1(\rho_0)}$. The system then
settles in an inhomogeneous steady state akin to MIPS that will be
investigated below.

\begin{figure}[t]
\centering\includegraphics[width=0.7\linewidth]{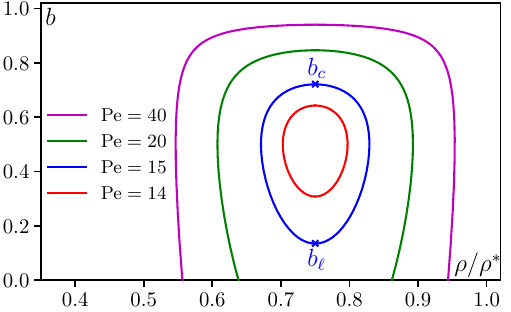}
\caption{Evolution of the spinodal lines with P\'eclet number $\Pe$ in
  the continuum theory for overdamped biased MIPS, as described in
  Sec.~\ref{sec:linear-stabl-OD}.}
\label{fig:spinodals}
\end{figure}

Interestingly, we find that in the fully biased case $b=1$, one has
$\eta_0=0$ so that $D_{\mathrm{eff}} = D>0$, leading to a linearly
stable homogeneous state at any density. There is thus a threshold
bias $b_c$ beyond which there is no phase separation. In the case of
the linear speed Eq.~(\ref{eq:u-simu}), we find that the two
spinodal lines marking the limit of linear stability have the form
\begin{equation}
  \label{eq:spinodals-u-lin} \rho=\frac{\rho^*}{4}\left[3 \pm
\sqrt{1-\frac{8 D}{u_0^2\eta_0}}\right].
\end{equation} The critical bias corresponds to the point where the
spinodals meet which is obtained, after replacing $\eta_0$ by its
expression Eq.~(\ref{eq:def:eta0}) in Eq.~(\ref{eq:spinodals-u-lin}),
as the solution of
\begin{equation}
  \label{eq:eq-bc}
\frac{1-b}{(2-b)^3}=\frac{2D\alpha}{u_0^2}\equiv\frac{2}{\Pe}
\end{equation} where in the last equality we have introduced a P\'eclet
number $\Pe=u_0^2/(\alpha D)$ as the ratio of the diffusion coefficient
due to active and passive motion. At large P\'eclet number,
Eq.~(\ref{eq:eq-bc}) gives a threshold bias $b_c=1-2/\Pe$. When
$\Pe<16$, we also get a lower threshold $b_\ell=-2+32/\Pe$, so that an
instability occurs only in the range $[b_\ell,b_c]$. Finally, for
$\Pe<27/2$, Eq.~(\ref{eq:eq-bc}) admits no physical solution and
homogeneous systems are stable at all values of the
bias. Fig.~\ref{fig:spinodals} recapitulates graphically the evolution
of the spinodal lines when varying $\Pe$.

The meeting of spinodal lines in a phase separation usually signal a
critical point. At $\Pe=16$, the lower threshold $b_\ell=0$ corresponds
to the critical point of unbiased MIPS which, in $d=2$, has been
measured to be in the Ising universality
class~\cite{maggi_universality_2021}, although with conflicting
reports~\cite{siebert_critical_2018}. Here we uncover a second
critical point at $b_c$ which may be destroyed by fluctuations in
$d=1$ but that we expect to survive in higher dimensions. Whether it
would be in the same universality class as the critical point of
unbiased MIPS is an interesting open problem.

\subsection{Phase separation}
\label{sec:phase-sep-OD}

As in MIPS, the instability identified in
Sec.~\ref{sec:linear-stabl-OD} signals a phase separation between
dense and dilute phases. However, a striking difference with MIPS is
that, because of the bias, the dense phase appears to be
moving. Indeed, as shown by Eq.~(\ref{eq:def:J0:mips}), there is a
particle current $J_0(\rho)=\ve\rho u(\rho)$ in a homogeneous phase at
density $\rho$ which needs not be the same in both phases. One can
then deduce the speed $c$ at which the dense phase is moving from mass
balance. Indeed, considering a fixed imaginary box around a front
connecting the gas phase at density $\rho_g$ and the liquid phase at
$\rho_\ell$, the inflow of mass is $J_0(\rho_g)$ and the outflow
$J_0(\rho_\ell)$. The front is thus moving at speed
\begin{equation}
  \label{eq:speed-ps}
  c=\frac{J_0(\rho_\ell)-J_0(\rho_g)}{\rho_\ell-\rho_g}.
\end{equation}

\begin{figure}[t]
\centering\includegraphics[width=1\linewidth]{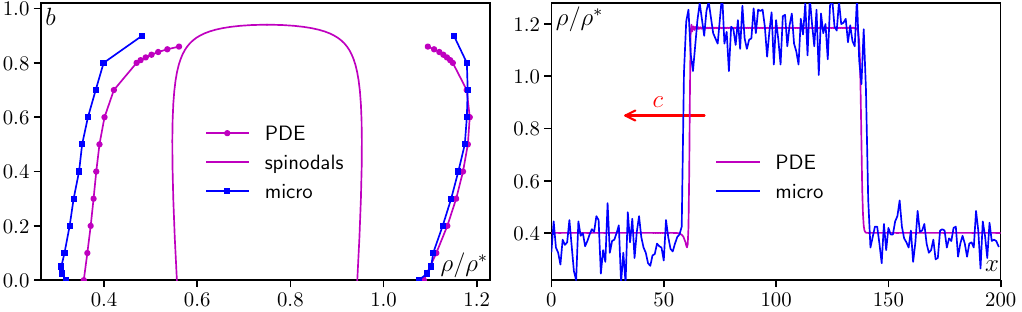}
\caption{Left: Phase diagram of the overdamped biased MIPS model
computed in simulations of the PDE (\ref{eq:evolrho:final:mips}) and
of the microscopic model Eq.~(\ref{eq:modelOD}). Right: instantaneous
density profiles for $b=0.6$. Parameters: $u_0=2$, $D=0.1$, $\alpha=1$
(giving $\Pe=40$), $\rho^*=200$, mean density $\rho_0=0.7
\rho^*$. $dt=0.005$ and $dx=0.5$ in integration of the PDE. $dt=0.1$
in integration of the microscopic model.}
\label{fig:binodals}
\end{figure}

In the unbiased case, the coexisting densities can be computed
analytically using effective thermodynamic
relations~\cite{solon_generalized_2018-1,solon_generalized_2018}. However,
when $b>0$, the additional gradient terms of even order in
Eq.~(\ref{eq:evolrho:final:mips}) invalidate this approach so that the
coexisting densities need to be determined numerically. We do this for
the linear velocity Eq.~(\ref{eq:u-simu}) both in numerical
integration of the field theory Eq.~(\ref{eq:evolrho:final:mips})
using a semi-spectral integration scheme with explicit time stepping
and in simulations of the microscopic model Eq.~(\ref{eq:modelOD})
using parallel updates and Euler time stepping. The result is shown as
a phase diagram in Fig.~\ref{fig:binodals} along with density
profiles. We see a good qualitative agreement between the microscopic
simulations and the continuum description although with quantitative
differences, especially in the gas binodal. In all our simulations, we
find that $\rho_\ell>\rho^*$ so that the particles are not moving in
the dense phase. The velocity $c$ of the dense phase given by
Eq.~(\ref{eq:speed-ps}) then reduces to
$c=-\rho_gu(\rho_g)/(\rho_\ell-\rho_g)$ and is thus always negative so
that the dense phase is moving upstream like traffic
jams~\cite{kerner_experimental_1998} or the freezed flocks of
colloidal rollers of Ref.~\cite{geyer_freezing_2019} which have a
similar phenomenology.

In Fig.~\ref{fig:binodals} (left), we see that the gas binodal
measured in the PDE appears to meet the spinodal before the critical
point, at a lower value of $b$. This is not a numerical artefact and
actually signals a transition from phase separation to microphase
separation at higher $b$ values. We will discuss this phenomenon in
more details in Sec.~\ref{sec:traffic} where this transition happens
further away from the critical point and is also observed in the
microscopic model.

Note that even if we observe phase-separated profiles in microscopic
simulations as shown in Fig.~\ref{fig:binodals} (right), we do not
expect it to be the asymptotic steady state in dimension
$d=1$. Indeed, as in the 1d Ising model, there is no surface tension
giving rise to a coarsening so that we expect domains of finite size
in the thermodynamic limit. Here, in order to observe large domains to
be able to measure binodals, we start from an initially
phase-separated state with different densities and use large values of
$\rho^*$ to effectively reduce fluctuations.

As we have seen, in the limit $b=1$ that is relevant to traffic flow,
homogeneous profiles are always stable, at odds with traffic
models~\cite{siebert_critical_2018,schreckenberg_discrete_1995} in
which, at high enough car density, a homogeneous car flow in
unstable, leading to traffic jam. As we show in the next section, to
recover this phenomenology, one needs to include inertial effects.

%%%%%%%%%%%%%%%%%%%%%%%%%%%%%%%%%%%%%%%%%%%%%%%%%%%%%%%%%%%%%%%%%%%%%%%

\section{Traffic-flow scenario: fully biased dynamics with inertia}
\label{sec:traffic}

\subsection{Definition of the model} We now consider an underdamped
version of the model of Sec.~\ref{sec:MIPS} in the fully biased $b=1$
case. It can be seen as a minimal model of traffic flow on a highway,
that is a long homogeneous portion of road with no crossing or traffic
light. A given car $i$ is characterized by its (one-dimensional)
position $x_i(t)$ at time $t$ on the road, and by its velocity
$v_i(t)=dx_i/dt$. The car $i$ has a preferred velocity $u(\rhot(x_i))$
that is computed in the same way as the quorum-sensing velocity of
Sec.~\ref{sec:MIPS}, using the kernel in Eq.~(\ref{eq:kernel}) to
compute the local density $\rhot(x_i)$. An important effect often
taken into account in the traffic flow modelling literature is the
fact that drivers take into account the density of cars in front of
them, but not the density of cars behind~\cite{schreckenberg_discrete_1995,chowdhury_statistical_2000}. Hence strictly speaking,
$\rhot(x_i)$ should be a measure of the density of cars just in front
of car $i$. For the sake of simplicity, we neglect this effect here
because it turns out to be inessential when discussing the analogy
with MIPS.

We assume that the velocity relaxes to the preferred speed $u(\rhot)$
with a relaxation rate $\gamma$ (and relaxation time
$\tau=\gamma^{-1}$), and is also subjected to an additive noise
$\xi_i(t)$,
\begin{equation}
  \label{eq:modelUD} \frac{dv_i}{dt} = - \gamma
\big[v_i-u\big(\rhot(x_i)\big)\big] + \gamma \sqrt{2D} \xi_i(t) .
\end{equation} The functional dependence of the target speed
$u(\rhot)$ is identical for all cars. In simulations we use the same
linear dependence Eq.~(\ref{eq:u-simu}) which reproduces the
non-monotonic variation of the car flow with density captured in the
``fundamental diagram of traffic flow''~\cite{siebel_fundamental_2006}. The white
noise $\xi_i(t)$ has the same correlation as in
Eq.~(\ref{eq:white:noise}). In the overdamped limit $\gamma\to \infty$, we exactly
recover the model of Sec.~\ref{sec:MIPS}  with $b=1$.

\subsection{Hydrodynamic equations} We introduce the single car
phase-space distribution $f(x,v,t)$, defined as the probability
density that a car with velocity $v$ is at position $x$ at time
$t$. Using the same local mean-field approximation as in
Sec.~\ref{sec:MIPS}, we assume that $f(x,v,t)$ is not correlated with
$\rhot$ so that the distribution $f(x,v,t)$ obeys the following
Fokker-Planck equation
\begin{equation} \label{eq:evolf} \partial_t f + \partial_x \big(
vf\big) -\gamma \partial_v \big[ \big(v-u(\rhot)\big) f\big] =
\gamma^2 D \partial_v^2 f.
\end{equation} The density field $\rho(x)$ and the car flux field
$w(x)$ are connected to the phase-space distribution $f(x,v,t)$
through the relations
\begin{equation} \label{eq:def:rho:w} \rho(x,t) =
\int_{-\infty}^{\infty} dv \, f(x,v,t), \qquad w(x,t) =
\int_{-\infty}^{\infty} dv \, v \, f(x,v,t).
\end{equation} In these notations, the local average velocity is equal
to $w/\rho$.  Integrating Eq.~(\ref{eq:evolf}) over the velocity $v$,
an evolution equation for $\rho$ (the continuity equation) is
obtained,
\begin{equation} \label{eq:evolrho} \partial_t \rho + \partial_x w =
0.
\end{equation} An evolution equation for $w$ is also needed. We
multiply Eq.~(\ref{eq:evolf}) by $v$ and then integrate it over $v$,
leading after integrations by part to:
\begin{equation} \label{eq:evolw} \partial_t w + \partial_x S + \gamma
\left( w -\rho u(\rhot) \right) = 0,
\end{equation} where we have introduced the second moment in $v$ of
the distribution $f$,
\begin{equation} \label{eq:def:S} S(x,t) = \int_{-\infty}^{\infty} v^2
f(x,v,t)\, dv.
\end{equation} Our goal is to expand Eq.~(\ref{eq:evolrho}) to
drift-diffusion order, that is, to second order in gradients. We thus
need to express $w$ in terms of $\rho$ to first order in gradients.
For time scales larger than $\gamma^{-1}$, we can neglect the time
derivative $\partial_t w$ in Eq.~(\ref{eq:evolw}) since $w$ is a fast
variable, and we get
\begin{equation} \label{eq:w:approx} w = \rho \ut - \frac{1}{\gamma}
\partial_x S,
\end{equation} with again $\ut=u(\rhot)$. We thus need to obtain $S$
to zeroth order in gradient. At this order, we get from
Eq.~(\ref{eq:evolf}) the evolution equation for $S$,
\begin{equation} \partial_t S + 2\gamma (S-\ut w) = 2\gamma^2 D\rho.
\end{equation} For time scales larger than $\tau=\gamma^{-1}$, we can
again neglect the time derivative, yielding
\begin{equation} S = \ut w+\gamma D\rho.
\end{equation} At the same order of approximation, we have from
Eq.~(\ref{eq:w:approx}) that $w=\ut\rho$, so that
\begin{equation} S = \ut^2\rho +\gamma D\rho.
\end{equation} Eq.~(\ref{eq:w:approx}) can then be rewritten to first
order in gradient as
\begin{equation} \label{eq:exprw} w = \rho \ut - \frac{1}{\gamma}
\partial_x \big( \rho \ut^2 \big) - D \partial_x \rho.
\end{equation} Replacing $w$ in Eq.~(\ref{eq:evolrho}) by its
expression (\ref{eq:exprw}), one eventually obtains a closed evolution
equation on the density field $\rho$,
\begin{equation} \label{eq:evolrho:final:traffic} \partial_t \rho =
-\partial_x \left( \rho \ut - 2\tau \rho \ut \partial_x \ut \right) +
\partial_x \left[ \left( \tau \ut^2 + D \right) \partial_x \rho
\right].
\end{equation}

We expand $\ut \equiv u(\rhot)$ as in Eq.~(\ref{eq:ut-exp}) to get a
fully explicit evolution equation for
$\rho$. Eq.~(\ref{eq:evolrho:final:traffic}) then takes the form
$\partial_t \rho = -\partial_x J$ with a particle current $J$ given by
\begin{equation}
  \label{eq:Jtraffic} J = J_0(\rho) - D_{\mathrm{eff}}(\rho)
\partial_x \rho + m(\rho) \partial_x^2 \rho + \Gamma_1 \partial_x^3
\rho - \Gamma_2 \partial_x \rho \,\partial_x^2 \rho
\end{equation} where $J_0(\rho)=\rho u$ as in
Eq.~(\ref{eq:def:J0:mips}), and the other coefficients are defined as
\begin{eqnarray}
\label{eq:def:Deff:traffic}
D_{\mathrm{eff}}(\rho) &=& D + \tau (\rho u^2)',\\
m(\rho) &=& \kappa \rho u',\\
\Gamma_1(\rho) &=& -2\tau \kappa \rho u u',\\
\Gamma_2(\rho) &=& 2\tau\kappa [\rho (uu''+u'^2)+ uu'].
\end{eqnarray}

\subsection{Linear instability of the homogeneous car flow} Traffic
congestion may be interpreted as resulting from the linear instability
of an homogeneous car flow.  We thus linearize
Eq.~(\ref{eq:evolrho:final:traffic}) around the homogeneous state of
density $\rho=\rho_0$, setting $\rho(x,t)=\rho_0+\delta\rho(x,t)$,
which yields to first order in $\delta \rho$
\begin{equation} \label{eq:evol:rho:lin} \partial_t \delta \rho = -
J_0'(\rho_0) \partial_x \delta \rho + D_{\mathrm{eff}}(\rho_0)
\partial_x^2 \delta \rho - m(\rho_0) \partial_x^3 \delta \rho -
\Gamma_1(\rho_0) \partial_x^4 \delta \rho.
\end{equation} The linear stability of the homogeneous state is
determined by the sign of the effective diffusion coefficient $D_{\rm
  eff}(\rho_0)$ defined in Eq.~(\ref{eq:def:Deff:traffic}).  When
$u(\rho_0)$ decreases fast enough, $D_{\rm eff}(\rho_0)$ can become
negative, signaling the instability of the homogeneous profile. In this case finite-wavelength perturbations grow
exponentially in time for $q<q^* =
\sqrt{|D_{\mathrm{eff}}(\rho_0)|/\Gamma_1(\rho_0)}$ (large $q$
perturbations are stabilized by the $q^4$ term).

The condition $D_{\mathrm{eff}}=0$ defines the spinodal line
$\gamma_s(\rho_0)$ with
\begin{equation}
  \label{eq:gammac} \gamma_s = -\frac{(\rho u^2)'(\rho_0)}{D}.
\end{equation}
For the linear velocity dependence Eq.~(\ref{eq:u-simu}),
Eq.~(\ref{eq:gammac}) gives a critical $\gamma_c=u_0^2/(3D)$ below
which the system is unstable in the density range between the spinodal
lines
\begin{equation}
  \label{eq:spinodals} \rho_s^{\pm}=\frac{\rho^*}{3}\left[2\pm
\sqrt{1-\frac{\gamma}{\gamma_c}}\right].
\end{equation}

\subsection{Phase separation}
\begin{figure}[t]
\centering\includegraphics[width=0.5\linewidth]{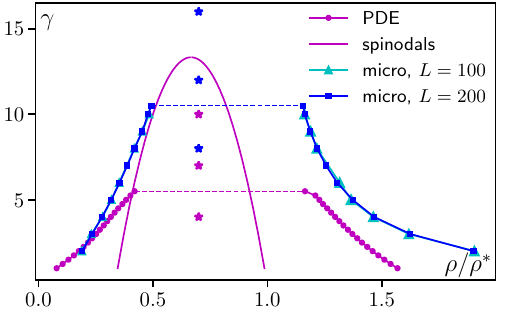}\\
\includegraphics[width=1\linewidth]{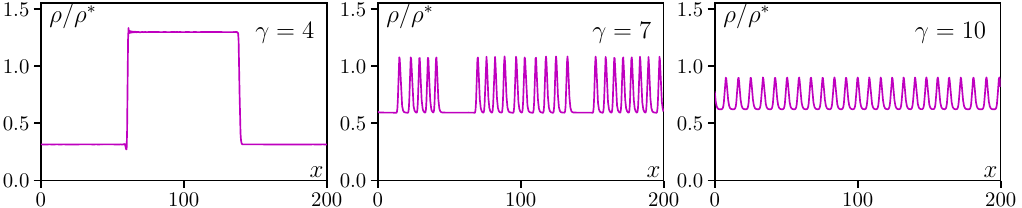}\\
\includegraphics[width=1\linewidth]{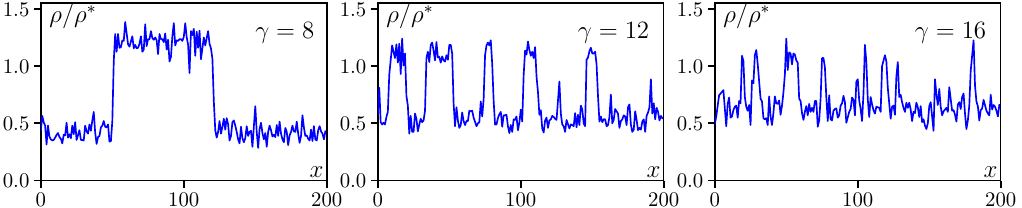}
\caption{Top: Phase diagram of the traffic model computed in
  simulations of the PDE (\ref{eq:evolrho:final:traffic}) and of the
  microscopic model Eq.~(\ref{eq:modelUD}). For the PDE, the spinodals
  correspond to the limit of linear stability of homogeneous profiles
  given by Eq.~(\ref{eq:spinodals}). The binodals are measured in the
  PDE and in the microscopic model for two different system sizes as
  the coexisting densities of the phase separated profiles which exist
  below the dashed lines.  The stars indicate the parameters of the
  snapshots shown below which were chosen to emphasize the
    qualitatively similar behaviour exhibited by the PDE and
    microscopic model. Middle row: Snapshots from numerical
  integration of the PDE starting from a phase separated initial
  condition. Bottom: Snapshots from simulations of the microscopic
  model starting from a phase separated initial condition. Parameters:
  $D=0.1$, $u_0=2$, $\rho^*=200$, $\rho_0=0.7 \rho^*$, system size
  $L=200$. $dt=0.005$ and $dx=0.5$ for the PDE. $dt=0.1$ for the
  microscopic model.}
\label{fig:binodals-traffic}
\end{figure}

Fig.~\ref{fig:binodals-traffic} shows the phase diagram in the
$\gamma-\rho$ plane as determined in the continuum theory
Eq.~(\ref{eq:evolrho:final:traffic}) and in the microscopic model. Let
us first discuss that obtained from the PDE.  It is markedly different
from the usual phase diagram of a liquid-gas transition. In
particular, we see that the lower binodal measured in simulations of
the PDE encounters the spinodal line at $\gamma\approx 5.5$. For
larger $\gamma$, phase separation is then impossible since the gas
phase would be unstable. As shown in the snapshots, one then observes
a microphase separation with an extensive number of high and low
density domains. The amplitude of the oscillation decreases as
$\gamma$ increases and vanishes at the critical point $\gamma_c$. Note
that for $\gamma=7$, the solution shown in
Fig.~\ref{fig:binodals-traffic} is not periodic but we expect that
adding a noise term to Eq.~(\ref{eq:evolrho:final:mips}) would allow
the periodic solutions to be reached, as is the case for the microphase
separation in flocking models~\cite{solon_pattern_2015}.

Compared to a passive liquid-gas phase separation or the usual MIPS,
the transition between phase and microphase separation is made
possible by the nonlinear drift term $J_0(\rho)$ in
Eq.~(\ref{eq:Jtraffic}). Indeed, this term does not affect the
spinodals that are controlled by $D_{\rm eff}$, but it does affect the
binodals. In contrast, for a conventional phase separation both the
binodals and spinodals are controlled by the free energy (or effective
free energy for MIPS~\cite{solon_generalized_2018-1}), and thus cannot
cross. The same physics is captured in a minimal setting by the
convective Cahn-Hilliard equation~\cite{golovin_convective_2001} in
which a convective non-linearity (that would enter in our notation as
$J_0\propto \rho^2$) is added to the standard Cahn-Hilliard equation.

 Comparing the phase diagrams from the PDE and microscopic model,
  we see that quantitatively the agreement is rather poor. As we argue
  in Sec.~\ref{sec:bridge}, it is in the fully-biased limit that our
  theory becomes more approximative, as can already be seen on
  Fig.~\ref{fig:binodals}. On the contrary, the effect of inertia is
  well captured as we will show in Sec.~\ref{sec:bridge}. Note that
  the discrepancy is not due to finite size effects, as seen in
  Fig.~\ref{fig:binodals-traffic} by comparing the phase diagram for
  two different system sizes. Nevertheless, despite the quantitative
differences, qualitatively, the same phenomenon is observed in the
microscopic model and the PDE: a phase separation at low values of
$\gamma$ crossing over to a microphase separation at larger $\gamma$
with domains of decreasing amplitude as $\gamma$ increases, as shown
in the snapshots of Fig.~\ref{fig:binodals-traffic} (bottom). The
  Supplementary Movie shows how the microphase separated state is
  reached from an initially phase-separated one in the microscopic
  model.

%%%%%%%%%%%%%%%%%%%%%%%%%%%%%%%%%%%%%%%%%%%%%%%%%%%%%%%%%%%%%%%%%%%%%%%%%%%%%%%%%%%%%%%%%

\section{Bridging MIPS and congested flow: inertial dynamics with tumbles}
\label{sec:bridge}

We have seen that the overdamped fully biased dynamics lead to a
stable homogeneous state. Phase separation is recovered either by
decreasing the bias (leading to MIPS) or by reintroducing inertia, with a
damping coefficient below a threshold value (leading to traffic
jams). In this perspective, MIPS and traffic jam seemingly appear as
disconnected phenomena, present in different parts of the phase
diagram of the model, and related to distinct physical
ingredients. However, a definite conclusion can only be reached by
considering the two-dimensional phase diagram in the parameter plane
$(b,\tau$), with $\tau=\gamma^{-1}$ the speed relaxation time, rather
than only the parameter lines $(b,\tau=0)$ [Section~\ref{sec:MIPS}]
and $(b=1,\tau)$ [Section~\ref{sec:traffic}] separately.

\subsection{Model with inertial dynamics and biased tumbling rates}

We combine the underdamped model of Sec.~\ref{sec:traffic} with the
biased tumbling dynamics of Sec.~\ref{sec:MIPS}.  The velocity $v_i$
evolves according to
\begin{equation}
  \frac{dv_i}{dt} = -\gamma \big[ v_i - u(\rhot) e_i(t) \big] + \gamma\sqrt{2D} \xi_i(t)
\end{equation}
where $e_i(t)$ has the same dynamics as in Sec.~\ref{sec:MIPS}, and
the white noise $\xi_i(t)$ has the correlation Eq.~(\ref{eq:white:noise}).

\subsection{Hydrodynamic description}
The derivation of the hydrodynamic equation follows the techniques of
Sec.~\ref{sec:MIPS} and~\ref{sec:traffic} combined. Let us introduce
the phase space densities $f_{+}(x,v,t)$ and $f_{-}(x,v,t)$ describing
the probability densities to find a particle at position $x$ with
velocity $v$ with $e_i=1$ or $e_i=-1$ respectively. The densities
$f_{+}$ and $f_{-}$ evolve according to
\begin{eqnarray}
 \!\!\!\!\!\!\!\! \partial_t f_{+} + \partial_x \big( vf_{+}\big) -\gamma \partial_v \left[ (v-\ut)f_{+}\right] &=& \gamma^2 D \partial_v^2 f_{+} -\alpha(1-b) f_{+} + \alpha f_{-},\\
 \!\!\!\!\!\!\!\! \partial_t f_{-} + \partial_x \big( vf_{-}\big) -\gamma \partial_v \left[(v+\ut)f_{-}\right] &=& \gamma^2 D \partial_v^2 f_{-} -\alpha f_{-} + \alpha (1-b) f_{+},
\end{eqnarray}
with again the shorthand notation $\ut = u(\rhot)$.  It is convenient
to introduce the total phase space density $f=f_{+}+f_{-}$, as well as
$g=f_{-}-(1-b) f_{+}$, by analogy with Sec.~\ref{sec:MIPS}. Keeping
the same notations, we have
\begin{eqnarray}
\label{eq:evol:f:mix} \partial_t f &+& \partial_x \big( vf\big) -\gamma \partial_v \left[ (v-\ve \ut)f\right] - (1+\ve) \gamma \partial_v (\ut g) = \gamma^2 D \partial_v^2 f,\\
\label{eq:evol:g:mix} \partial_t g &+& \partial_x \big( vg\big) -\gamma \partial_v \left[ (v+\ve \ut)g\right] - \mu \gamma \partial_v (\ut f) = \gamma^2 D \partial_v^2 g - \tilde{\alpha} g,
\end{eqnarray}
with $\mu=2(1-b)/(2-b)$ and $\tilde{\alpha}=\alpha (2-b)$.  We use the
fields $\rho(x,t)$, $w(x,t)$ and $S(x,t)$ defined as the first moments
in $v$ of the phase space distribution $f(x,v,t)$ as in
Eqs.~(\ref{eq:def:rho:w}) and (\ref{eq:def:S}). Similarly, we also
introduce the following auxiliary fields related to the function
$g(x,v,t)$:
\begin{eqnarray}
  q(x,t) &=& \int_{-\infty}^{\infty} g(x,v,t)\, dv,\\
\bar{w}(x,t) &=& \int_{-\infty}^{\infty} v\, g(x,v,t)\, dv,\\
\bar{S}(x,t) &=& \int_{-\infty}^{\infty} v^2 g(x,v,t)\, dv.
\end{eqnarray}
Integrating Eq.~(\ref{eq:evol:f:mix}) over $v$, one finds
the evolution equation for $\rho$, which reads
\begin{equation}
  \label{eq:rho:mix:v0} \partial_t \rho = -\partial_x w,
\end{equation}
so that we need to evaluate $w$ at first order in gradients.
We apply a similar method as in the previous cases, by determining the evolution equation for
the successive low-order moments of $f$ and $g$. Typically, the evolution equation of the
moment of order $n$ involves a gradient of the moment of order $n+1$, successively leading
to truncations at lower and lower order in the gradient expansion.
In addition, moments of order $n>0$ can be considered as fast variables and their time derivative
can be neglected after a coarse-graining in time.
These approximations allow us to truncate and close the moment hierarchy at order $n=2$.
A detailed derivation is given in \ref{sec:appendix}.
We eventually obtain an explicit expression for $w$,
\begin{equation} \label{eq:w:inertial:tumble}
  w = \big[ \ve \ut - \eta (\ut\partial_x \ut)\big] \rho - \big[ \tilde{\eta} \ut^2 + D\big] \partial_x \rho,
\end{equation}
with
\begin{eqnarray}
  \zeta &=& \frac{4(1-b)\beta\gamma}{(2-b)^2},\\
  \eta &=& \frac{\zeta}{\tilde{\alpha}} + \frac{2}{\gamma}(\ve^2+\zeta),\\
  \tilde{\eta} &=& \frac{\zeta}{\tilde{\alpha}} + \frac{1}{\gamma}(\ve^2+\zeta).
\end{eqnarray}
From Eq.~(\ref{eq:rho:mix:v0}), we finally get a closed evolution
equation for $\rho$,
\begin{equation}
  \label{eq:evolrho:final:bridge} \partial_t \rho = -\partial_x \big(\ve \rho \ut - \eta \rho \ut \partial_x \ut\big) + \partial_x \left[ \big( \tilde{\eta} \ut^2 + D \big) \partial_x \rho \right].
\end{equation}
A more explicit expression of $\eta$ and $\tilde{\eta}$ is obtained by
expanding these coefficients to leading order in the relaxation time
$\tau = \gamma^{-1}$, for $\tau\to 0$:
\begin{eqnarray}
  \eta &=& \frac{4(1-b)}{\alpha (2-b)^3} + 2\tau\, \frac{1+(1-b)^2}{(2-b)^2},\\
  \tilde{\eta} &=& \frac{4(1-b)}{\alpha (2-b)^3} + \tau \, \frac{b^2}{(2-b)^2}.
\end{eqnarray}
These expressions match the results of the previous
sections. For $\tau=0$ (overdamped dynamics), one recovers
$\eta=\eta'=\eta_0$ in agreement with Eq.~(\ref{eq:def:eta0}), while
for $b=1$, one recovers $\eta=2\tau$ and $\eta'=\tau$ consistently
with Eq.~(\ref{eq:evolrho:final:traffic}).

We expand $\ut \equiv u(\rhot)$ as in Eq.~(\ref{eq:ut-exp}) to get a
fully explicit evolution equation for
$\rho$. Eq.~(\ref{eq:evolrho:final:bridge}) then takes the form
$\partial_t \rho = -\partial_x J$ with a particle current $J$ given by
\begin{equation}
  \label{eq:Jbridge} J = J_0(\rho) - D_{\mathrm{eff}}(\rho) \partial_x \rho + m(\rho) \partial_x^2 \rho + \Gamma_1(\rho) \partial_x^3 \rho - \Gamma_2(\rho) \partial_x \rho \,\partial_x^2 \rho
\end{equation}
with again the same current $J_0(\rho)=\ve \rho u(\rho)$, and where
\begin{eqnarray}
  D_{\mathrm{eff}}(\rho) &=& D+ \eta \rho u u' + \tilde{\eta} u^2,\\
  m(\rho) &=& \kappa \ve \rho u',\\
  \Gamma_1(\rho) &=& -\eta\kappa \rho u u',\\
  \Gamma_2(\rho) &=& \eta\kappa\rho (uu''+u'^2)+2\tilde{\eta}\kappa uu'.
\end{eqnarray}
% Eq.~(\ref{eq:Jbridge}) is consistent with previous results as one
% recovers Eq.~(\ref{eq:Jmips}) in the limit $\tau\to 0$ and
% Eq.~(\ref{eq:Jtraffic}) for $b=1$.

\begin{figure}[t]
\centering\includegraphics[width=0.6\linewidth]{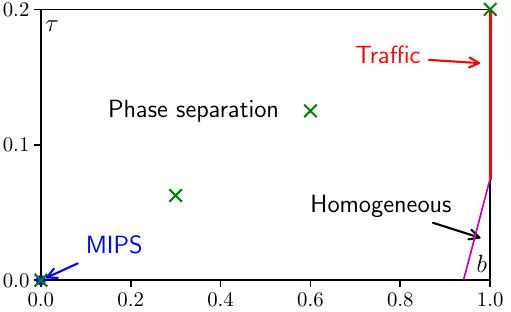}
\centering\includegraphics[width=1\linewidth]{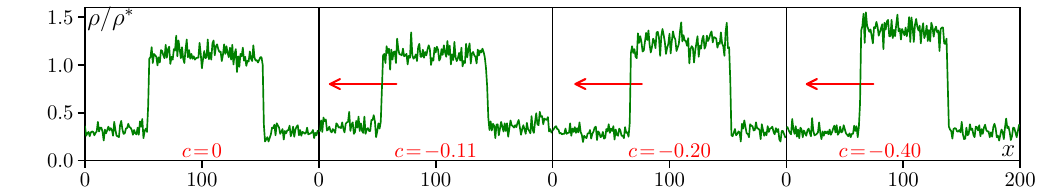}
\caption{Phase diagram in the $(b,\tau)$-plane of the bias $b$ and
  relaxation time $\tau$ for $\Pe=40$. The purple line delimit
  the regions where only homogeneous states are observed at high $b$
  and small $\tau$ and the region where phase separation is observed
  for a range of densities.  The green crosses indicate the position
  of the snapshots shown in the lower panel [from left to right
  ($b=0$, $\tau=0$), ($b=0.3$, $\tau=1/16$), ($b=0.6$, $\tau=1/8$),
  ($b=1$, $\tau=1/5$)]. They show that one can continuously change
  parameters from the unbiased MIPS (blue dot) to the (micro) phase
  separation seen in the fully-biased traffic model of
  Sec.~\ref{sec:traffic}. The red arrows emphasize the motion of the
  dense phase at the speed $c$, measured
    numerically. Parameters: system size $L=200$, $\rho^*=200$,
  $\rho_0=0.7\rho^*$, $u_0=2$, $D=0.1$.}
\label{fig:phase:diag}
\end{figure}

\subsection{Linear stability analysis}

We linearize Eq.~(\ref{eq:evolrho:final:bridge}) around the
homogeneous state of density $\rho=\rho_0$, setting
$\rho(x,t)=\rho_0+\delta\rho(x,t)$, which yields to first order in
$\delta \rho$
\begin{equation}
  \label{eq:evol:rho:lin:bridge} \partial_t \delta \rho = - J_0'(\rho_0) \partial_x \delta \rho + D_{\mathrm{eff}}(\rho_0) \partial_x^2 \delta \rho -m(\rho_0) \partial_x^3 \delta \rho - \Gamma_1(\rho_0) \partial_x^4 \delta \rho,
\end{equation}
As before the homogeneous state may become linearly unstable when
$D_{\mathrm{eff}} < 0$, in which case the instability occurs for small wave-vectors
$q<q^* = \sqrt{|D_{\mathrm{eff}}(\rho_0)|/\Gamma_1(\rho_0)}$.

In Fig.~\ref{fig:phase:diag}, we plot the phase diagram in the
$(b,\tau)$ plane computed from the condition $D_{\mathrm{eff}} < 0$
for a fixed P\'eclet number $\Pe=u_0^2/(\alpha D)=40$. The purple line
separates a region at small $\tau$ and large $b$ in which homogeneous
profiles are stable at any density (and phase separation is thus never
observed) from a region where phase separation is observed is some
density range. From that plot, it is clear that the phase separation
seen in unbiased MIPS ($b=0$, $\tau=0$) is continuously connected to
the traffic congestion ($b=1$, $\tau>0$). The snapshots shown in
Fig.~\ref{fig:phase:diag} (bottom) confirm that, indeed, such a
continuous interpolation also exists in the microscopic model.

\begin{figure}
\centering\includegraphics[width=0.6\linewidth]{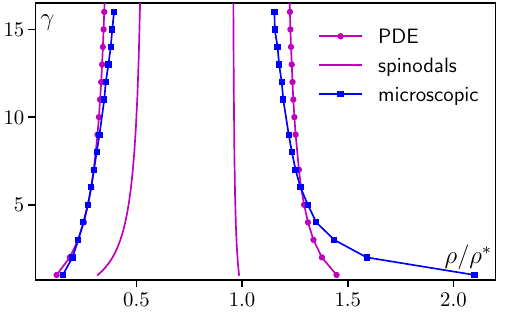}
\caption{Phase diagram of the model with inertia and biased tumbles
  with $b=0.6$ compared with the hydrodynamic theory
  Eq.~(\ref{eq:Jbridge}). For these parameter values, only phase
  separation is observed between the binodals, microphase separation
  happening only at larger $b$ values. Parameters: $D=0.1$, $u_0=2$,
  $\rho^*=200$, $\rho_0=0.7 \rho^*$, system size $L=200$; $dt=0.005$
  and $dx=0.5$ for the PDE; $dt=0.1$ for the microscopic model.}
\label{fig:phase:diag-b}
\end{figure}

To test the validity of the continuum theory
  Eq.~(\ref{eq:Jbridge}) taking into account the effect of both
  inertia and tumbles, we show in Fig.~\ref{fig:phase:diag-b} a
  comparison of the phase diagrams obtained at $b=0.6$ for the PDE and
  for the microscopic model. We see that the quantitative agreement is
  much better than in the case $b=1$ shown in
  Fig.~\ref{fig:binodals-traffic}, with large discrepancies only at
  small $\gamma$, as expected because of the small $\tau$ ({\it i.e.}
  large $\gamma$) approximation used to derive the continuum
  equation. The discrepancy seen in Fig.~\ref{fig:binodals-traffic}
  is thus expected to come from the limit $b\to 1$, which is indeed singular with
  several coefficients vanishing in Eq.~(\ref{eq:Jmips}), rather than
  from the effect of inertia.

\begin{figure}[t]
\centering\includegraphics[width=0.6\linewidth]{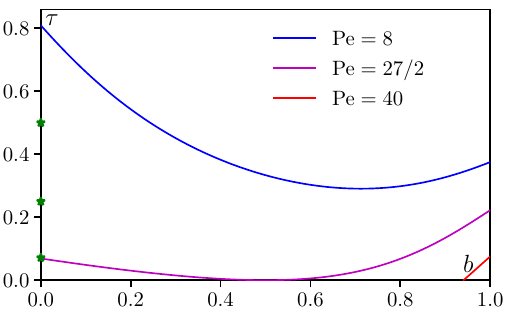}
\includegraphics[width=1\linewidth]{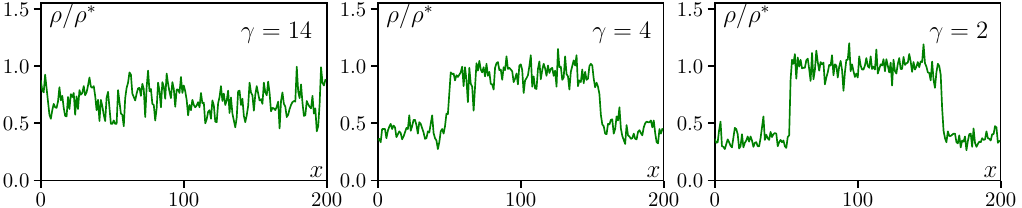}
\caption{Top: Phase diagram in the $(b,\tau)$-plane, as in
  Fig.~\ref{fig:phase:diag}, for several values of
  $\Pe=u_0^2/(\alpha D)$. The green stars indicate the position of the
  snapshots shown below. Bottom: Snapshots from the microscopic model
  in the unbiased case $b=0$ at $\Pe=16$ ($u_0=2$, $\alpha=1$, $D=0.25$)
  showing that MIPS can be observed only when inertia is strong enough
  ({\it i.e.} large $\tau$/small $\gamma$).}
\label{fig:phase:diag-D}
\end{figure}

It is interesting to see how the phase diagram evolves with P\'eclet
number. As shown in Fig.~\ref{fig:phase:diag-D}, we see that at lower
activity it becomes non-monotonous. Below the critical value
$\Pe=27/2$ already identified in Sec.~\ref{sec:linear-stabl-OD},
inertia is necessary to observe phase separation. More generally we
find that inertia always has a destabilizing effect on homogeneous
profiles and thus promotes MIPS.  The snapshots in
Fig.~\ref{fig:phase:diag-D} confirm that, even for $b=0$, this
stabilization of MIPS by inertia also happens in the microscopic
model. In contrast, for self-propelled disks, a strong enough
inertia was found to destroy the phase
separation~\cite{mandal_motility-induced_2019}. This can be traced
back to the details of collisions of self-propelled disks. At large
inertia, colliding particles tend to bounce back rather than stall as in
the overdamped limit, thus suppressing the slowdown leading to
MIPS~\cite{mandal_motility-induced_2019}. This effect is absent in our
model where collisions are absent (interactions proceed through the dependence
of particle speed of the local density),
and where inertia acts as a delay in the sensing of the local density.

\section{Conclusion}
In this paper we have introduced and studied a one-dimensional model
featuring run-and-tumble particles with tumbles that are biased in one
direction. They interact via a target speed $u(\rho)$ that is a
decreasing function of the local density and to which the particles
relax at a finite rate $\gamma$. In Sec.~\ref{sec:MIPS}, we first
considered the case of overdamped particles with biased tumbles which
could describe systems showing MIPS in presence of an external bias
such as chemotaxis toward a food source for bacteria. We then
considered in Sec.~\ref{sec:traffic} the completely biased overdamped
case which is most naturally interpreted as a traffic model on a
highway showing traffic jams. Finally, in Sec.~\ref{sec:bridge} we
have put both ingredients together to look at the phase diagram in
terms of both the bias and relaxation rate. In the three cases, we have
derived the hydrodynamic equation for the density field and used it to
study the linear stability of homogeneous profiles. We then compared
the phase diagrams obtained with numerical simulations of the microscopic
model and of the continuum equations.

Let us recapitulate the most salient features observed in our
model. (i) We find that the ``usual'' (overdamped unbiased) MIPS
survives the addition of bias, with an ordered phase moving against
the flow of particles. However, MIPS is lost at any activity level at
large values of the bias. (ii) Inertia promotes phase separation and
restores it in the fully biased case. Even in the unbiased case, MIPS can
be observed in presence of inertia for activity levels such that it is
absent in the overdamped case. (iii) The full model allows us to show
that MIPS and the traffic jams seen in the fully biased model are
essentially the same phenomenon as one can go continuously from one to
the other. (iv) For some parameter values, we observe that
inhomogeneous profiles transition from phase separation to microphase
separation. This happens when a binodal line crosses the spinodal
line so that a macroscopic phase coexistence becomes unstable. Such a
phenomenon, studied before in the context of a convective
Cahn-Hilliard equation~\cite{golovin_convective_2001}, is made
possible here by the bias which generates terms of odd order in gradient
in the continuous description.

In the microscopic model, the study of the phase to microphase
separation transition is complicated by the fact that, in one
dimension, fluctuations are already expected to break a macroscopic
phase separation into an extensive number of domains. We circumvented
this difficulty by considering large densities in order to reduce
fluctuations. However, we expect our results to extend to higher
dimensions, as long as the bias remains along one preferred direction,
in which case the transition will be more easily studied. Whether this
transition can happen for any value of the bias or only beyond a
critical bias remains an open question left for future work. In
addition, considering the higher dimensional case would allow one to assess
the universality class of the critical point of the phase (or
microphase) separation at non-zero bias. Finally, our model builds on
the quorum-sensing model of MIPS which is well described by the active
model B~\cite{wittkowski_scalar_2014}, showing macroscopic phase
separation and a positive surface tension. It could be interesting to
study a biased version of self-propelled disks which fall in the class
of active model B+~\cite{tjhung_cluster_2018} showing negative surface
tension~\cite{bialke_negative_2015,solon_generalized_2018} and a
bubbly phase
separation~\cite{caporusso_motility-induced_2020,shi_self-organized_2020}. A
potential additional instability of the phase separation due to the
bias, as we observe, could lead to a very rich phenomenology.

\subsubsection*{Acknowledgements.}

We thank Hugues Chaté for useful discussions.  

\appendix

\section{Evaluation of $w$ for the inertial dynamics with tumbles}
\label{sec:appendix}

We evaluate the field $w$ using a gradient expansion to second order of the evolution equations
for the low-order moments of $f$ and $g$.
Multiplying Eq.~(\ref{eq:evol:f:mix}) by $v$ and
integrating over $v$, we get after an integration by part
\begin{equation}
  \partial_t w + \partial_x S + \gamma (w-\ve \ut \rho) + (1+\ve)\gamma \ut q = 0.
\end{equation}
On time scales much larger than $\tau = \gamma^{-1}$, we can neglect $\partial_t w$, leading to
\begin{equation}
  \label{eq:w:mix:v1} w = \ve \ut \rho - (1+\ve) \ut q-\frac{1}{\gamma} \partial_x S.
\end{equation}
To obtain $w$, we thus need to determine $q$ to first order in
gradient, and $S$ to zeroth order in gradient. Integrating
Eq.~(\ref{eq:evol:g:mix}) over $v$, we get
\begin{equation}
  \label{eq:q:mix:v0} \partial_t q + \partial_x \bar{w} = -\tilde{\alpha} q.
\end{equation}
On time scales much larger than $\alpha^{-1}$, we thus get
\begin{equation} \label{eq:q:mix:v1} q = -\frac{1}{\tilde{\alpha}}
\partial_x \, \bar{w}.
\end{equation}
It follows that we need to determine $\bar{w}$ at zeroth order in
gradient.  Multiplying Eq.~(\ref{eq:evol:g:mix}) by $v$ and
integrating over $v$, we get after an integration by part,
\begin{equation}
  \label{eq:wt:mix:v0} \partial_t \bar{w} + \partial_x \bar{S} + \gamma \ve \ut q + (1+\ve) (1-b) \gamma \ut \rho = -(\gamma+\tilde{\alpha}) \bar{w}.
\end{equation}
Neglecting time and space derivatives and taking into account
Eq.~(\ref{eq:q:mix:v1}) to neglect the term proportional to $q$, the
expression of $\bar{w}$ simplified to
\begin{equation}
  \label{eq:wt:mix:v1} \bar{w} = -(1+\ve)(1-b) \beta\gamma\, \ut \rho
\end{equation}
where we have introduced $\beta = (\gamma+\tilde{\alpha})^{-1}$. It
follows from Eqs.~(\ref{eq:q:mix:v1}) and (\ref{eq:wt:mix:v1}) that
\begin{equation}
  \label{eq:q:mix:v2} q = (1+\ve) \frac{(1-b)\beta\gamma}{\tilde{\alpha}} \, \partial_x (\ut \rho).
\end{equation}
We now determine $S$ to zeroth order in gradient. Multiplying
Eq.~(\ref{eq:evol:f:mix}) by $v^2$ and integrating over $v$, we get
after integration by part and neglected gradient terms
\begin{equation}
  \partial_t S + 2\gamma (S-\ve \ut w) + 2\gamma (1+\ve) \, \ut \bar{w} = 2\gamma^2 D\rho.
\end{equation}
Neglecting the time derivative $\partial_t S$ since $S$ is a fast
field and using respectively the expressions (\ref{eq:w:mix:v1}) and
(\ref{eq:wt:mix:v1}) of $w$ and $\bar{w}$ to zeroth order in gradient,
we get
\begin{equation}
  \label{eq:S:mix:v1} S = (\ve^2 + \zeta) \ut^2 \rho + \gamma D\rho
\end{equation}
with
\begin{equation}
  \zeta = \frac{4(1-b)\beta\gamma}{(2-b)^2}.
\end{equation}
Plugging the expressions (\ref{eq:q:mix:v2}) of $q$ and
(\ref{eq:S:mix:v1}) of $S$ into Eq.~(\ref{eq:w:mix:v1}), 
we finally obtain Eq.~(\ref{eq:w:inertial:tumble}) for the expression of $w$.

%%%%%%%%%%%%%%%%%%%%%%%%%%%%%%%%%%%%%%%%%%

\bigskip
\bibliographystyle{plain_url}
\bibliography{traffic-flow}

\end{document}